\documentclass{IEEEtran}
\usepackage{mathrsfs}
\usepackage{svg}
\usepackage[ruled,linesnumbered]{algorithm2e}
\usepackage{verbatim}
\usepackage{soul}
\ifCLASSINFOpdf
\else
\fi
\usepackage[cmex10]{amsmath}
\usepackage{amsthm,cite,url,graphicx,booktabs,lipsum,color,bm,soul}
\usepackage{pifont,tikz,paralist,multirow,amssymb}

\theoremstyle{definition}

\makeatletter
\newcommand{\removelatexerror}{\let\@latex@error\@gobble}
\makeatother

\begin{document}
\title{Latency Aware Semi-synchronous Client Selection and Model Aggregation for Wireless Federated Learning}

\author{Liangkun~Yu,~\IEEEmembership{Graduate~Student~Member,~IEEE,} Xiang~Sun,~\IEEEmembership{Member,~IEEE,}
Rana~Albelaihi,~\IEEEmembership{Graduate~Student~Member,~IEEE,}
and Chen~Yi,~\IEEEmembership{Member,~IEEE} 
\thanks{L. Yu, X. Sun, and R. Albelaihi are with the SECNet Lab., University of New Mexico, Albuquerque, NM 87131, USA. E-mail: $\{$liangkun, sunxiang, ralbelaihi$\}$@unm.edu.\par
C. Yi is with the Chongqing Key Laboratory of Signal and Information Processing, School of Communication and Information Engineering, Chongqing University of Posts and Telecommunications, Chongqing 400065, China. E-mail: yichen@cqupt.edu.cn.\par
This work was supported by the National Science Foundation under Award CNS-2148178.\par
This work has been submitted to the IEEE for possible publication. Copyright may be transferred without notice, after which this version may no longer be accessible.
}
}

\maketitle

\begin{abstract}
Federated learning (FL) is a collaborative machine learning framework that requires different clients (e.g., Internet of Things devices) to participate in the machine learning model training process by training and uploading their local models to an FL server in each global iteration. Upon receiving the local models from all the clients, the FL server generates a global model by aggregating the received local models. This traditional FL process may suffer from the straggler problem in heterogeneous client settings, where the FL server has to wait for slow clients to upload their local models in each global iteration, thus increasing the overall training time. One of the solutions is to set up a deadline and only the clients that can upload their local models before the deadline would be selected in the FL process. This solution may lead to a slow convergence rate and global model overfitting issues due to the limited client selection. In this paper, we propose the \ul{L}atency awar\ul{E} \ul{S}emi-synchronous client \ul{S}election and m\ul{O}del aggregation for federated lear\ul{N}ing (LESSON) method that allows all the clients to participate in the whole FL process but with different frequencies. That is, faster clients would be scheduled to upload their models more frequently than slow clients, thus resolving the straggler problem and accelerating the convergence speed, while avoiding model overfitting. Also, LESSON is capable of adjusting the tradeoff between the model accuracy and convergence rate by varying the deadline. Extensive simulations have been conducted to compare the performance of LESSON with the other two baseline methods, i.e., FedAvg and FedCS. The simulation results demonstrate that LESSON achieves faster convergence speed than FedAvg and FedCS, and higher model accuracy than FedCS. 
\end{abstract}

\begin{IEEEkeywords}
Federated learning, client selection, model aggregation, semi-synchronous
\end{IEEEkeywords}
\IEEEpeerreviewmaketitle

\section{Introduction}
With the development of Internet of Things (IoT), numerous smart devices, such as smartphones, smartwatches, and virtual reality headsets, are widely used to digitize people's daily lives. Traditionally, a huge volume of data generated by these IoT devices is uploaded to and analyzed by a centralized data center that generates high-level knowledge and provides corresponding services to users, thus facilitating their lives \cite{Sun:2016:EdgeIoT}. A typical example is smart home, where various IoT devices, such as smart meters, thermostats, motion detectors, and humidity sensors, are deployed to monitor the status of the smart homes. The data generated by the IoT devices would be uploaded to a centralized data center, which applies deep reinforcement learning model to intelligently and autonomously control, for example, the smart bulbs and air conditioners in the smart homes, to improve the quality of experience and reduce the energy usage of smart homes \cite{8675180}.

\begin{figure}[!htb]
	\centering	
	\includegraphics[width=0.8\columnwidth]{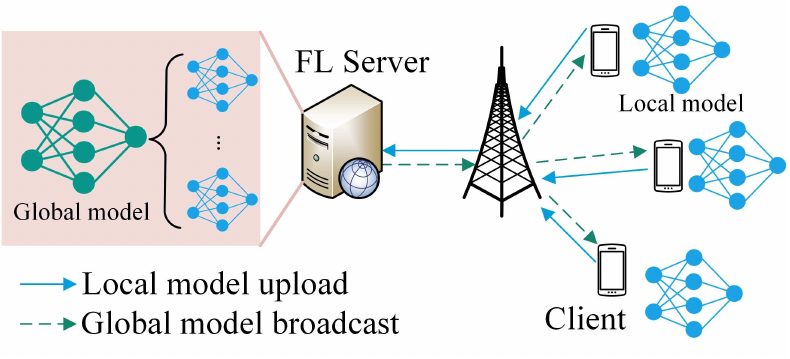} 
	\caption{Wireless federated learning.}
	\label{fig:wfl} 
\end{figure}

On the other hand, sharing data with third-party data centers may raise privacy concerns as data generated by the IoT devices may contain personal information, such as users' locations and personal preferences \cite{DHANVIJAY2019113,Granjal:2015:SIoT}. As a result, various policies have been made, such as General Data Protection Regulation (GDPR) made by the European Union \cite{goddard2017eu}, to regulate and hinder data sharing. In order to fully utilize these personal data while preserving privacy (i.e., without sharing the data), federated learning (FL) is proposed to distributively train machine learning models by enabling different IoT devices to analyze their data locally without uploading them to a central facility \cite{mcmahan2017communication}. A typical example to use FL is to train a next word prediction model, which is used to predict what word comes next based on the existing text information \cite{hard2018federated}. Basically, as shown in Fig. \ref{fig:wfl}, an FL server would first initialize the parameters of the global model and then broadcast the global model to all the clients via wireless networks. Each client would train the received global model based on its local data sets (i.e., their text messages) and upload the updated model to the FL server via wireless networks. The FL server then aggregates the received models from the clients to generate a new global model, and then starts a new iteration by broadcasting the new global model to the clients. The iteration continues until the model is converged.     

The traditional FL method can resolve the privacy issue to allow the clients to train the model locally, and it applies the synchronous strategy, where the FL server has to wait until it receives the models from all the clients in each global iteration. This may lead to the straggler problem when the configurations of clients are heterogeneous, meaning that they have different computing and communications capabilities. Hence, some stragglers take much longer time to train and upload their models in a global iteration because of their lower computing and communications capabilities, and thus significantly prolong the model training process. To resolve the straggler problem, many client selection methods have been proposed \cite{imteaj2020fedar,wu2020safa,reisizadeh2020straggler,xu2019elfish,Rana:2021:APS}, which would select the qualified clients that can finish their model training and uploading before a predefined deadline. Normally, client selection and resource allocation are jointly optimized to maximize the number of the selected qualified clients. Selecting the qualified clients can resolve the straggler problem, but may raise other issues. First, the proposed client selection may significantly reduce the number of participated clients, which may slow down the convergence speed \cite{wang2018cooperative}, thus leading to longer training latency (which equals to the sum of the latency for all the global iterations). Second, the proposed client selection may result in the model overfitting issue caused by the reduction of data diversity. That is, if the FL server only selects the qualified clients to participate in the model training, then the generated model can only fit the data samples in these qualified clients, but not the non-qualified clients. The model overfitting issue would be compounded if the data samples of the qualified clients are not sufficient \cite{shorten2019survey}. The other solution to solve the straggler problem is to apply asynchronous FL, where the FL server does not need to wait until the deadline expires for each global iteration, but would update the global model once it receives a local model from a client \cite{li2020survey}. However, the asynchronous FL may suffer from 1) the high communications cost since both the FL server and clients will more frequently exchange their models, and 2) the stale issue, where some slow clients are training based on an outdated global model, which may lead to slow convergence rate or even global model divergence \cite{xu2021asynchronous,Damaskinos:2020:Fleet}.  

In order to solve the slow convergence and model overfitting issues in the synchronous FL while avoiding model divergence in the asynchronous FL, we propose a semi-synchronous FL method, i.e., \ul{L}atency awar\ul{E} \ul{S}emi-synchronous client \ul{S}election and m\ul{O}del aggregation for federated lear\ul{N}ing (LESSON). The basic idea of LESSON is to allow all the clients to participate in the whole learning process with different frequencies. Specifically, the clients are clustered into different tiers based on their model training and uploading latency. The clients in a lower tier (i.e., lower model training and uploading latency) would participate in the learning process more frequently than those in a higher tier (with higher model training and uploading latency). As a result, the straggler problem can be resolved (since the FL server does not need to wait for stragglers in each global iteration) and the model overfitting problem can be fixed (since all the clients join the learning process to provide high data diversity). The main contributions of the paper are summarized as follows.
\begin{enumerate}
    \item In LESSON, we propose a latency-aware client clustering method to group different clients into different tiers based on their model training and uploading latency.   
    \item In LESSON, we design a model aggregation based on the proposed client clustering. The designed model aggregation method  determines the weight and time for the clients in each tier to aggregate their local models. Also, LESSON can flexibly adjust its performance to change the tradeoff between model accuracy and model convergence rate. 
    \item The performance of LESSON is evaluated via extensive experiments. The results demonstrate that LESSON achieves a faster convergence speed than FedAvg and FedCS, and higher model accuracy than FedCS.
\end{enumerate}

The rest of this paper is organized as follows. The related work is summarized in Section II. System models are described in Section III. Section IV elaborates the proposed LESSON algorithm, which comprises client clustering and model aggregation. In Section V, the performance of LESSON is compared with the other two baseline algorithms via extensive simulations, and simulation results are analyzed. Finally, Section VI concludes this paper.

\section{Related Work} 
How to solve the straggler issue is one of the main challenges in synchronous FL. The existing solutions mainly focus on jointly optimizing client selection and resource allocation. Nishio and Yonetani \cite{Nishio} aimed to maximize the number of selected clients that can finish their model training and uploading before a predefined deadline in each global iteration. By assuming that the selected clients have to iteratively upload their models to the FL server, they designed FedCS that jointly optimizes the uploading schedule and client selection to achieve the objective. Abdulrahman and Tout \cite{FedMCCS} designed a similar client selection method FedMCCS. The goal of FedMCCS is to maximize the number of selected clients, who can not only finish the  model training and uploading before a predefined deadline but also guarantee the resource utilization less than the threshold to avoid device dropout. Albelaihi \emph{et al.} \cite{Rana:2021:APS} proposed a client selection method that tries to achieve the same objective as FedCS, but they argued that the latency of a client in waiting for the wireless channel to be available for model uploading should be considered; otherwise, the selected clients may not upload their local models before the deadline. Yu \emph{et al.} \cite{9509751} proposed to dynamically adjust and optimize the tradeoff between maximizing the number of selected clients and minimizing the total energy consumption of the selected clients by picking suitable clients and allocating appropriate resources (in terms of CPU frequency and transmission power) in each global iteration. Shi \emph{et al.} \cite{9149138} also jointly optimized the client selection and resources allocation for FL. However, the objective is to minimize the overall learning latency (which equals the product of the average latency of one global iteration and the number of global iterations), while achieving a certain model accuracy. They formulated a system model to estimate the number of global iterations given the global model accuracy requirement. All the mentioned client selection method for synchronous FL can alleviate the straggler problem, but may lead to the slow convergence rate and model overfitting issues as we illustrated previously. 

Instead of selecting qualified clients to avoid stragglers, Li \emph{et al.} \cite{MLSYS2020_38af8613} proposed to let fast clients train their local models by running more gradient descent iterations in each global iteration. In this way, the fast and slow clients could upload their models at the similar time, but the fast clients may provide better models to fit their local data samples, thus potentially speedup the model convergence. This method, however, may lead to local model overfitting issue when some fast clients run too many gradient descent iterations over the limited data samples. The overfitted local models would significantly slow down the global model convergence. Wu \emph{et al.}\cite{wu2021fedadapt} proposed to perform spilt learning, where a global neural network is divided into two parts. The parameters in the former and latter layers are trained in the clients and the FL server, respectively. As a result, the computational complexity of the clients is reduced, thus potentially reducing the training time and energy consumption of the clients. 

Other works aim to design asynchronous FL, where the FL server does not need to wait for the selected clients to upload their local models; instead, once the FL server receives a local model from a client, it would aggregate the received local model to update the global model, and then send the updated global model to the client \cite{chen2020asynchronous,lu2019differentially,gu2020privacy,lian2018asynchronous}. However, as mentioned before, the fast clients and the FL server have to more frequently exchange their models, thus leading to higher communications cost for both the fast clients and FL server \cite{chai2020fedat,feyzmahdavian2014delayed}. Meanwhile, in asynchronous FL, the slower clients may train their local models based on an outdated global model, which results in slow convergence or even leads to model divergence \cite{10.1145/3035918.3035933,Wei:2016:SAD}.

To overcome the drawbacks in synchronous and asynchronous FL, we propose the semi-synchronous FL to allow all the clients to participate in the whole learning process with different frequencies. Although the term "semi-synchronous FL" has been used by the existing works, the definitions are different from what we defined in this paper. For example, Stripelis and Ambite \cite{stripelis2021semi} defined semi-synchronous FL as the clients to train their local models over different sizes of local data sets, depending on their computing capabilities. The semi-synchronous FL proposed in \cite{hao2020time} periodically re-selects a number of clients, and follows the same method as asynchronous FL to aggregate and update the global model.      



\section{System Models}
\subsection{Federated learning preliminary}
The idea of FL is to enable distributed clients to cooperatively train a global model such that the global loss function, denoted as ${\bm{\mathcal{F}}\left( \bm{\omega} \right)}$, can be minimized. That is,

\begin{equation}
\mathop { \arg \min }\limits_{\bm{\omega}} {\bm{\mathcal{F}}\left( \bm{\omega} \right)} = \mathop { \arg \min }\limits_{\bm{\omega}} \sum\limits_{i \in \bm{\mathcal{I}}} \frac{{\left| {\bm{\mathcal{D}}_i} \right|}}{{\left| {\bm{\mathcal{D}}} \right|}} {{f_i}\left( \bm{\omega} \right)}, 
\label{fl_obj}
\end{equation}
where $\bm{\omega}$ is the set of the parameters for the global model, $\bm{\mathcal{I}}$ is the set of the selected clients, ${\left| {\bm{\mathcal{D}}} \right|}$ is the number of the training data samples of all the clients, ${\left| {\bm{\mathcal{D}}_i} \right|}$ is the number of the training data samples at client $i$ (where $\bm{\mathcal{D}} = \bigcup\limits_{i \in \bm{\mathcal{I}}}  \bm{\mathcal{D}}_i  $), and ${f_i}\left( \bm{\omega} \right)$ is the local loss function of client $i$, i.e., 
\begin{equation}
    {f_i}\left( \bm{\omega}  \right)=\frac{1}{{\left| {{{\bm{\mathcal{D}}_i}}} \right|}}\sum\limits_{n \in {{\bm{\mathcal{D}}_i}}} {f\left( {\bm{\omega} ,{\bm{a}_{i,n}},{b_{i,n}}} \right)}.
\end{equation}
Here, $\left( {{\bm{a}_{i,n}},{b_{i,n}}} \right)$ is the input-output pair for the $n^{th}$ data sample in user $i$'s data set, and $f\left( {\bm{\omega} ,{\bm{a}_{i,n}},{b_{i,n}}} \right)$ captures the error of the local model (with parameter $\bm{\omega}$) over $\left( {{\bm{a}_{i,n}},{b_{i,n}}} \right)$.  


In each global iteration, FL comprises four steps. 
\begin{enumerate}
\item In the $k$-th global iteration, the FL server broadcasts the global model generated in the previous global iteration, denoted as ${{\bm{\omega}}^{\left( {k-1} \right)}} $, to all the selected clients.

\item  Each client $i$ trains its local model over its local data set ${\bm{\mathcal{D}}_i}$, i.e., $\bm{\omega}_i^{\left( {k } \right)} = \bm{\omega} _i^{\left( k-1 \right)} - \delta \nabla {f_i}\left( {\bm{\omega} _i^{\left( k -1\right)}} \right)$, where $\delta$ is the learning rate.

\item After deriving the local model ${\bm{\omega} _i^{\left( k \right)}}$, client $i$ uploads its local model to the FL server.

\item The FL server aggregates the local models from the clients and updates the global model based on, for example, FedAvg \cite{mcmahan2017communication}, i.e., 
${{\bm{\omega}}^{\left( {k } \right)}} = \sum\limits_{i \in \bm{\mathcal{I}}} {\frac{{\left| {\bm{\mathcal{D}}_i} \right|}}{{\left| {\bm{\mathcal{D}}} \right|}}{\bm{\omega}} _i^{\left( {k } \right)}} $. 
\end{enumerate}
The global iteration keeps executed to update the global model ${{\bm{\omega}}^{\left( {k } \right)}}$ until the global model converges.

\begin{figure*}[!htb]
	\centering	
	\includegraphics[width=1.8\columnwidth]{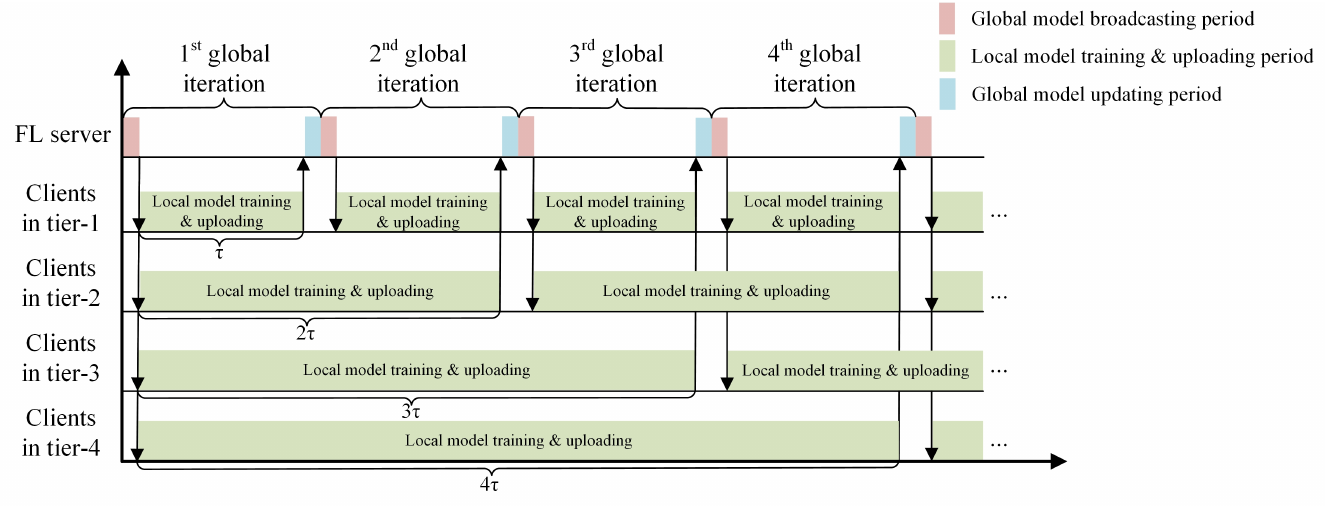} 
	\caption{Illustration of client scheduling in LESSON.}
	\label{fig:LESSON_sch} 
\end{figure*}

\subsection{Latency models of a client}
There are four steps in each global iteration for FL, and so the latency of a global iteration equals the sum of the latency among these four steps. The local model training latency in Step 2) and local model uploading latency in Step 3) are different among the clients, depending on their computing and communications capacities. In addition, the broadcast latency in Step 1) and model aggression latency in Step 4) are the same for all the clients, and they are negligible as compared to local model training and uploading latency. Thus, we define the latency of client $i$ in a global iteration as follows.  
\begin{equation}
    t_i=t_i^{comp}+t_i^{upload},
\label{eq:client_latency}
\end{equation}
where $t_i^{comp}$ is the computing latency of client $i$ in training its local model over its local data samples, and $t_i^{upload}$ is the uploading latency of client $i$ in uploading its local model to the FL server. 
\subsubsection{Computing latency}
The computing latency of client $i$ in a global iteration can be estimated by \cite{yang2020delay}
\begin{equation} 
t^{comp}_i =\theta\log_2\left(\frac{1}{\epsilon}\right)\frac{C_i  {\left| {\bm{\mathcal{D}}}_i \right|}} {f_i}, 
\label{eq:comp_latency}
\end{equation}
where $\theta$ is a constant determined by the structure of the desired model; $\theta\log_2\left(\frac{1}{\epsilon}\right)$ is the number of iterations required for a desired accuracy $\epsilon$; $C_i$ is the number of CPU cycles required for training one data sample of the model ${\bm{\omega}}$, varied by the complexity of a sample and CPU design;  ${\left| {\bm{\mathcal{D}}}_i \right|}$ is number of samples on each client; and $f_i$ is CPU frequency for training the model.

\subsubsection{Uploading latency}
The achievable data rate of client $i$ can be estimated by
\begin{equation}
{r_i} = b\log_2\left( {1 + \frac{{{p_i}g_i}}{{{N_0}}}} \right),
\end{equation}
where $b$ is the amount of bandwidth allocated to each participate client, ${p_i}$ is the transmission power of client $i$, ${g_i}$ is the channel gain from client $i$ to the BS, and ${N_0}$ is the average background noise and inter-cell interference power density. Assume that the size of the local model is $s$, and so the latency of client $i$ in uploading its local model to the BS is
\begin{equation} 
t^{upload}_i =\frac{s}{r_i}=\frac{s}{b\log_2\left(1+\frac{pg_i}{N_0}\right)}.
\label{eq_upload_latency}
\end{equation}

\section{\ul{L}atency awar\ul{E} \ul{S}emi-synchronous client \ul{S}election and m\ul{O}del aggregation for federated lear\ul{N}ing (LESSON)}
Different from synchronous FL, the proposed LESSON method aims to allow all the clients to participate in the whole learning process while avoiding the straggler problem. The basic idea of LESSON is to cluster the clients into different tiers based on their latency and the deadline. The clients in different tiers would train and upload their local models at different frequencies.    

\subsection{Latency-aware client clustering}
Denote $\tau$ as the deadline of a global iteration. The FL server would accept all the models uploaded from the clients before the deadline $\tau$ and reject the rest in each global iteration. Hence, we cluster the clients into a number of tiers, and $x_{ij}$ is used to indicate client $i$ is in tier $j$ (i.e., $x_{ij}=1$) or not (i.e., $x_{ij}=0$). Basically, if client $i$ can finish its local model training and uploading before deadline $\tau$, i.e., $t_i \le \tau$, then client $i$ is in tier-1, i.e., $x_{i1}=1$. Similarly, if client $i$ can finish its local model training and uploading between $ \tau$ and $2 \times \tau$, i.e., $\tau< t_i \le 2 \times \tau$, then client $i$ is in tier-2, i.e., $x_{i2}=1$. The following equation provides a general mathematical expression to cluster client $i$ into a specific tier.    


\begin{align}
x_{ij} = \left \{ {\begin{array}{*{20}{l}}
{1,}&{{\rm{if  }}\ \tau\times (j-1)<t_i\le\tau\times j,}\\
{0,}&{{\rm{otherwise,}}}
\end{array}} \right.
\label{eq:client_cluster}
\end{align}
where $j$ is the index of tiers. 

\subsection{Semi-synchronized model aggregation}
In each global iteration, clients from different tiers would upload their local models, and the FL server is able to estimate when a client may upload its local model according to its associated tier. Fig. \ref{fig:LESSON_sch} provides one example to illustrate the scheduling of the clients from four tiers in LESSON. For example, the clients in tier-1 are expected to upload their local models in each global iteration, and the clients in tier-2 are expected to upload their local models in every two global iterations. Denote $k$ as the index of the global iterations, and let $y_{jk}$ be the binary variable to indicate whether the clients in tier-$j$ are expected to upload their local models by the end of $k^{th}$ global iteration, where    
\begin{align}
y_{jk} = \left \{ {\begin{array}{*{20}{l}}
{1,}&{{\rm{if  }}\ k\%j=0,}\\
{0,}&{{\rm{otherwise,}}}
\end{array}} \right. 
\end{align}
where $\%$ is the modulo operation, and so $k\%j=0$ indicates $k$ is divisible by $j$. Meanwhile, let $z_{ik}$ be the binary variable to indicate whether client $i$ is expected to upload its local model by the end of $k^{th}$ global iteration ($z_{ik}=1$) or not ($z_{ik}=0$) , where
\begin{equation}
    z_{ik}=x_{ij}y_{jk}.
\end{equation}
Based on the value of $ z_{ik}$, the FL server would expect which clients will upload their local models in global iteration $k$, and then aggregate all the received local models based on
\begin{equation}
    {{\bm{\omega}}^{\left( {k } \right)}} = \sum\limits_{i \in \bm{\mathcal{I}}} {\frac{{\left| {\bm{\mathcal{D}}_i} \right|}}{\sum\limits_{i \in \bm{\mathcal{I}}}{\left| {\bm{\mathcal{D}}_i} \right|}z_{ik}}{\bm{\omega}} _i^{\left( {k } \right)}z_{ik}},
\label{eq:global_model_update}
\end{equation}
where $\sum\limits_{i \in \bm{\mathcal{I}}}{\left| {\bm{\mathcal{D}}_i} \right|}z_{ik}$ indicates the total number of the data samples among all the clients. who would upload their local models in global iteration $k$.  

Note that, in synchronous FL (e.g., FedAvg), each selected client would update its local model $\bm{\omega}_i^{\left( {k } \right)}$ based on Eq. \eqref{eq:model_update_syn} in each global iteration.  
\begin{equation}
    \bm{\omega}_i^{\left( {k } \right)} = \bm{\omega} _i^{\left( k-1 \right)} - \delta \nabla {f_i}\left( {\bm{\omega} _i^{\left( k-1 \right)}} \right),
\label{eq:model_update_syn}
\end{equation}
where $\delta$ is the step size, which is the same for all the selected clients. In LESSON, different clients update their local models in different frequencies, depending on their associated tiers, and so it is reasonable to adopt different step sizes for the clients in different tiers \cite{chen2020asynchronous}. Here, we adjust the step size of a client proportional to its tier (i.e., $\delta \times j$). Thus, when $k\%j=0$, client $i$ in tier-$j$ would update its local model based on
\begin{equation}
    \bm{\omega}_i^{\left( {k } \right)} = \bm{\omega} _i^{\left( k -j\right)} - j\delta \nabla {f_i}\left( {\bm{\omega} _i^{\left( k-j \right)}} \right)
\label{eq:local_model_update}
\end{equation}

\subsection{Summary of LESSON}
Algorithm \ref{alg:LESSON} summarizes the LESSON algorithm. Initially, the FL server estimates the latency of all the clients and cluster the clients into different tiers based on Eq. \eqref{eq:client_cluster}, i.e., Steps 1-2 in Algorithm \ref{alg:LESSON}. Then, the FL server broadcasts the initial global model ${\bm{\omega}}^ {\left( 0\right)}$ to start the collaborative model training process, which comprises many global iterations. In any global iteration $k$, each client trains and updates its local model ${\bm{\omega}}_{i}^ {\left( k\right)}$ based on Eq. \eqref{eq:local_model_update}. If $z_{ik}=1$, client $i$ should upload its local model to the FL server by the end of global iteration $k$. Then, client $i$ would wait until it receives the updated global model ${\bm{\omega}}^ {\left( k\right)}$ from the FL server to start the next round of local model training. On the other hand, the FL server keeps receiving the local models from the clients in global iteration $k$. Once the deadline expires, the FL server updates the global model ${\bm{\omega}}^ {\left( k\right)}$ based on Eq. \eqref{eq:global_model_update}, and then broadcasts the new global model ${\bm{\omega}}^ {\left( k\right)}$ to the clients, who just uploaded their local models in global iteration $k$. 

Note that the deadline of a global iteration, i.e., $\tau$, is a very important parameter to adjust the performance of LESSON. Specifically, if $\tau \rightarrow +\infty$, there is only one tier and all the clients in the network would be in this tier, and so LESSON acts as FedAvg, where the FL server has to wait until it receives the local models from all the clients in each global iteration. On the other hand, if $\tau \rightarrow 0$, LESSON acts as asynchronous FL, where clients with different latency (i.e., $t_i$) will be in different tiers. The FL server will immediately aggregate the local models from the clients with the same latency and then update and broadcast the global model. Thus, one of the advantages of LESSON is that it can change $\tau$ to adjust its performance, and we will discuss how $\tau$ affects the performance of LESSON in Section V.     

\begin{figure}[!t]
 \removelatexerror
\begin{algorithm}[H]
\label{alg:LESSON}
\SetAlgoLined
\caption{LESSON algorithm}
    Estimate the latency of all the clients based on Eq. \eqref{eq:client_latency}.
    
    Cluster the clients into different tiers based on Eq. \eqref{eq:client_cluster}.
    
    The FL server initializes the global model ${\bm{\omega}}^ {\left( 0\right)}$ and broadcasts to all the clients.
    
    $k=1$.

    \For{each global iteration $k$}{
    
    \SetKwBlock{Fna}{\textnormal{\textbf{Client side}:\{}}{}
    \Fna{
    
    Update its local model based on Eq. \eqref{eq:local_model_update};
    
    \If{$z_{ik}=1$}{Upload its local model ${\bm{\omega}}_{i}^ {\left( k\right)}$ to the FL server;
    
    Wait until it receives the global model ${\bm{\omega}}^ {\left( k\right)}$ broadcast by the FL server; }
    
    $k:=k+1$;}\}\
    
    \SetKwBlock{Fnb}{\textnormal{\textbf{FL server side}:\{}}{}
    \Fnb{
    Receive all the local models uploaded from the clients during global iteration $k$;
    
    Update the global model ${\bm{\omega}}^ {\left( k\right)}$ based on Eq. \eqref{eq:global_model_update};
    
    Broadcast the updated global model ${\bm{\omega}}^ {\left( k\right)}$ to the corresponding clients; 
    
    $k:=k+1$;}\}\
    }
\end{algorithm}
\end{figure}

%
%
%
%
%
%
%
%
%
%
%
%
%
%
%
%
%
%

\section{Simulation}
In this section, we conduct extensive simulations to evaluate the performance of LESSON.  
\begin{figure}[!htb]
	\centering	
	\includegraphics[width=0.8\columnwidth]{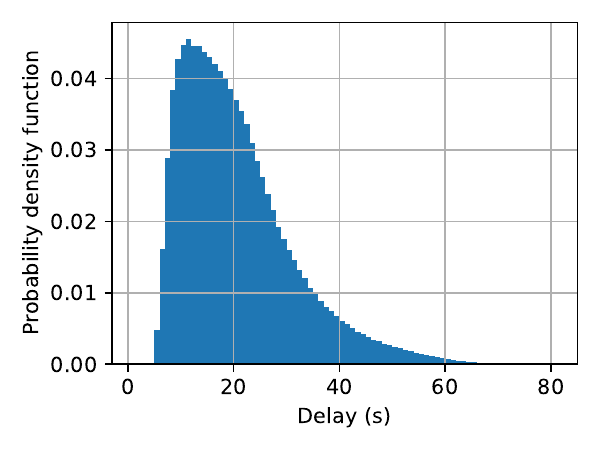} 
	\caption{Clients' latency distribution.}
	\label{fig:cld} 
\end{figure}

\begin{figure}[!htb]
	\centering	
	\includegraphics[width=0.8\columnwidth]{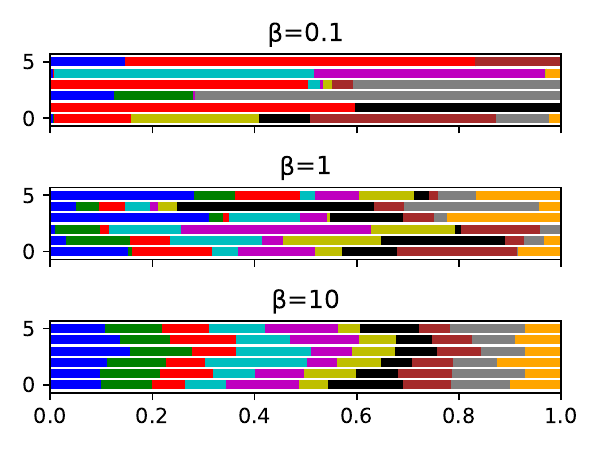} 
	\caption{Probability distribution of 10 categories samples for 5 clients with different $\beta$. }
	\label{fig:noniid} 
\end{figure}

\subsection{Simulation setup}
\subsubsection{Configuration of clients} 
We assume that there are $50$ clients that are uniformly distributed in a 2 km $\!\!\times\!\!$ 2 km area, which is covered by a BS located at the center of the area. All the clients upload their local models to the FL server via the BS. The pathloss between the BS and client $i$ is calculated based on $128.1+37.6d_i$, where $d_i$ is the distance in kilometer between the BS and client $i$. Meanwhile, the transmission power $p_i$ is set to be $1$ Watt for all the clients, and the amount of available bandwidth for each client in uploading its local model is $30$ kHz. In addition, each client has around ${\left| {\bm{\mathcal{D}}}_i \right|}=1,000$ data samples (i.e., $50,000$ combined for all clients) to train its local model. The number of CPU cycles required for training one data sample (i.e., $C_i$) among clients are randomly selected from a uniform distribution, i.e., $C_i \sim \mathcal{U}\left(3,5\right)\!\times\!10^5$ CPU cycles/sample. The CPU frequency of a client $f_i$ is also randomly selected from a uniform distribution, i.e., $f_i \sim \mathcal{U}\left(0.8,3\right)$ GHz. Other simulation parameters are listed in Table \ref{table_1}.

\begin{table}[!htb]
	\centering
	\caption{Simulation parameters}
	\label{table_1}
	\begin{tabular}{ll}
		\toprule
		\textbf{Parameter} & \textbf{Value} \\
		\midrule
	    Noise and inter-cell interference ($N_0$) & $-94$ dBm\\
	    Size of the local model ($s$)& $100$ kbit\\
	    Number of local iterations $(\theta\log_2(\frac{1}{\epsilon}))$ & $1\!\times\!\log_2(\frac{1}{0.05})$\\
	   Number of local samples ${\left| {\bm{\mathcal{D}}}_i \right|}$  &  $1000$\\
	   Number of local epochs  & $1$\\
	   Number of local batch size  & $20$\\
	   Non-IID Dirichlet distribution parameter $\beta$   & $1$\\
		\bottomrule
	\end{tabular}%
	\label{tab:sim_para}%
\end{table}%

Fig. \ref{fig:cld} shows the probability density function of the latency (i.e., $t_i$) among all the clients in a specific time instance. The median value in the clients' latency is around 10 seconds, and so we initially set up the deadline $\tau$ of a global iteration to be 10 seconds, and we will change the value of $\tau$, later on, to see how it affects the performance of LESSON.       
\subsubsection{Machine learning model and training datasets}
We will use two benchmark datasets to train the corresponding machine learning model.
\begin{enumerate}
    \item CIFAR-10\cite{Krizhevsky2009LearningML} is an image classification dataset containing 10 labels/classes of images, each of which has 6,000 images. 
    Among the 60,000 images, 50,000 are used for model training and 10,000 for model testing.

    \item MNIST \cite{lecun1998gradient} is a handwritten digits dataset that includes many $28\times28$ pixel grayscale images of handwritten single digits between 0 and 9. The whole dataset has a training set of 60,000 examples and a test set of 10,000 examples.
\end{enumerate}

We apply the convolutional neural network (CNN) 
to classify the CIFAR-10 images. The CNN model has four 3x3 convolution layers (where the first layer has 32 channels, and each of the following three layers has 64 channels. Also, only the first two layers are followed with $2\times2$ max pooling), followed with a dropout layer with rate of 75$\%$, a fully connected 256 units ReLu layer, and a 10 units softmax output layer. There are totally 1,144,650 parameters in this CNN model.

With respect to MNIST, which is a much simpler image dataset than CIFAR-10, a smaller CNN model has been used. Specifically, the CNN model has two 5x5 convolution layers (where the two layers have 6 and 16 channels, respectively, each of which is followed with a $2\times2$ max pooling), followed with two fully connected ReLu layers with 120 and 84 units, respectively, and a 10 units softmax output layer. There are totally 61,706 parameters in this CNN model.

In addition, we partition the MNIST/CIFAR-10 dataset among the 50 clients based on non-independent and identical distribution (Non-IID), and the probability of having $\eta_m$ images in label class $m$ at client $i$ is assumed to follow a Dirichlet distribution \cite{li2021federated}, i.e.,

\begin{equation}
    f(\eta_1,\eta_2,\dots,\eta_M;\beta)=\frac{\Gamma(\beta M)}{\Gamma(\beta)^M}\prod_{m=1}^{M}\eta_m^{\beta-1},
\end{equation}
where $M$ is the total number of label classes (i.e., $M=10$ for both CIFAR-10 and MNIST), $\Gamma()$ is the gamma function\footnote{$\Gamma(z)=\int_{0}^{\infty }x^{z-1}e^{-x}dx$}, and $\beta$ is the concentration parameter that determines the level of label imbalance. A larger $\beta$ results in more balanced data partition among different labels within a client (i.e., much closer to IID), and vice versa. Fig. \ref{fig:noniid} shows how different labels of images are distributed by varying $\beta$. In addition, we assume each client would train its CNN model over ${\left| {\bm{\mathcal{D}}}_i \right|}=20$ data samples locally based on stochastic gradient descent (SGD) with base learning rate $\delta=0.1$ \cite{MAL-083} and epoch equal to 1.    
\textcolor{blue}{\subsubsection{Baseline comparison methods} }

\subsubsection{Baseline algorithms} 
The performance of LESSON will be compared with the other two baseline client selection algorithms, i.e., FedCS \cite{Nishio} and FedAvg \cite{wang2018cooperative}. FedCS only selects the clients that can finish their model uploading before the deadline $\tau$ in each global iteration, i.e., only the clients in tier-1 will be selected to participate in the training process. In FedAvg, all the clients in the network will be selected to participate in the training process for each global iteration. That is, the FL server would wait until it receives the local models from all the clients, and then update the global model for the next global iteration.    

\subsection{Simulation results}
\begin{figure*}[!htb]
	\centering	
	\includegraphics[width=\textwidth]{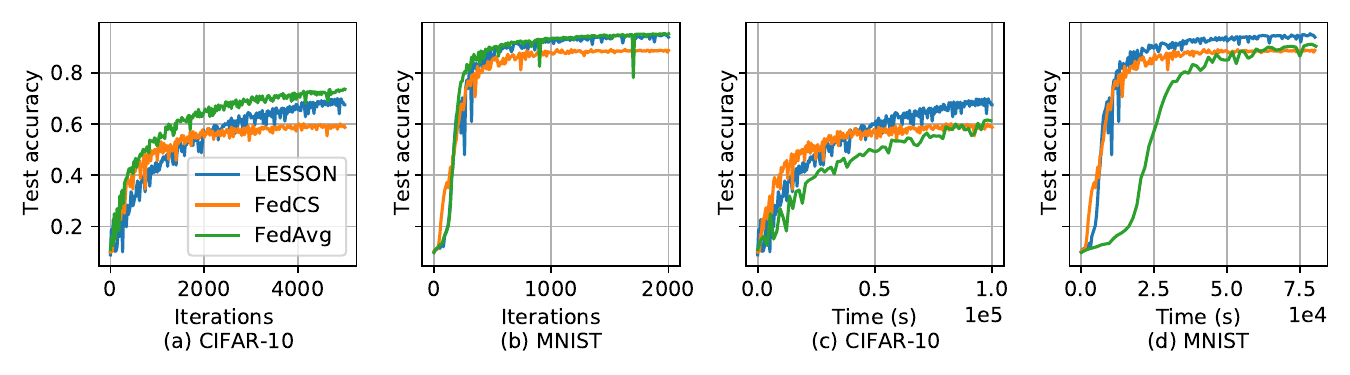} 
	\caption{Test accuracy of different algorithms for CIFAR-10 and MNIST, where (a) test accuracy vs number of global iterations for CIFAR-10, (b) test accuracy vs number of global iterations for MNIST, (c) test accuracy vs time in CIFAR-10, and (d) test accuracy vs time in MNIST.}
	\label{fig:var_data} 
\end{figure*}

Assume that $\beta=1$ and $\tau=20$ seconds. Fig. \ref{fig:var_data} shows the test accuracy of the three algorithms over the global iterations and simulation time for CIFAR-10 and MNIST. From Figs \ref{fig:var_data}(a) and \ref{fig:var_data}(b), we can find that LESSON and FedAvg have similar test accuracy, i.e., $\sim 70\%$ for CIFAR-10 and $\sim 95\%$ for MNIST. However, the test accuracy achieved by FedCS is lower than LESSON and FedAvg, i.e., $\sim 60\%$ for CIFAR-10 and $\sim 90\%$ for MNIST. This is because FedCS only selects fast clients to participate in the training process, and so the derived global model can only well fit the data samples for fast clients, not slow clients, thus reducing the model accuracy. Meanwhile, the convergence rate with respect to the number of global iterations for LESSON and FedAvg is also very similar, which is slightly faster than FedCS. However, by evaluating the convergence rate with respect to the time, as shown in Figs \ref{fig:var_data}(c) and \ref{fig:var_data}(d), we find out that LESSON is faster than FedAvg. For example, the global model in LESSON has already converged at 20,000 seconds for MNIST, but the global model in FedAvg is still under-trained. This is because the FL server in FedAvg has to wait until the local models from all the clients have been received in each global iteration, and thus the latency of a global iteration incurred by FedAvg is much higher than that incurred by LESSON. Table \ref{tab:delay_iteration} shows the average delay of a global iteration incurred by different algorithms, where the average latency of a global iteration incurred by LESSON is 48 seconds faster than FedAvg. 
As a result, FedAvg only runs around $588$ global iterations at 20,000 seconds in Figs and \ref{fig:var_data}(d), respectively, but LESSON runs $2,000$ global iterations.

\begin{table}[!htb]
	\centering
	\caption{Average latency per global iteration for different algorithms.}

	\begin{tabular}{cc}
		\toprule
		\textbf{Algorithms} & \textbf{Average latency of a global iteration}  \\
		\midrule
	   FedAvg & 68 seconds\\
	   FedCS & 20 seconds\\
	   LESSON & 20 seconds\\
		\bottomrule
	\end{tabular}%
	\label{tab:delay_iteration}%
\end{table}%


\begin{figure*}[!htb]
	\centering	
	\includegraphics[width=\textwidth]{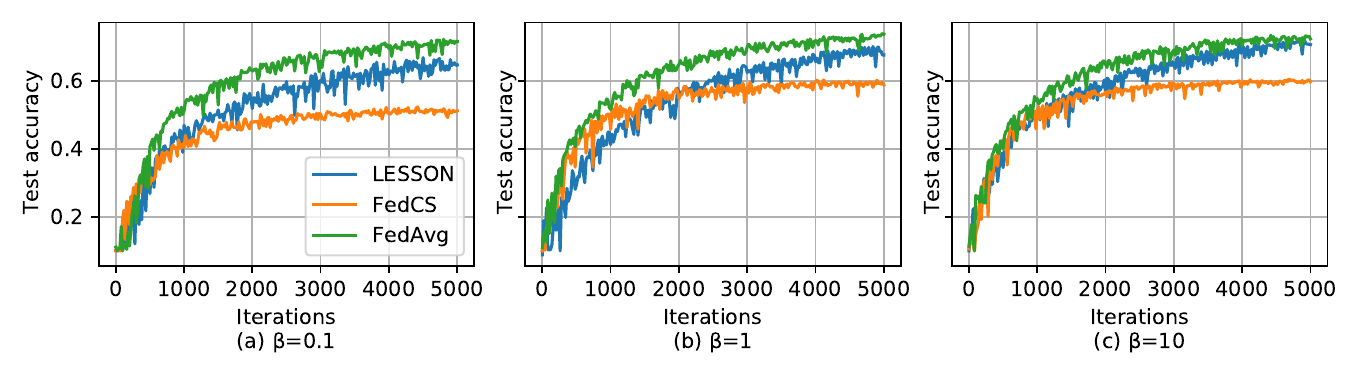} 
	\caption{Test accuracy over number of global iterations for CIFAR-10, where (a) $\beta=0.1$, (b) $\beta=1.0$, and (c) $\beta=10.0$.}
	\label{fig:var_beta_iteration} 
\end{figure*}
\begin{figure*}[!htb]
	\centering	
	\includegraphics[width=\textwidth]{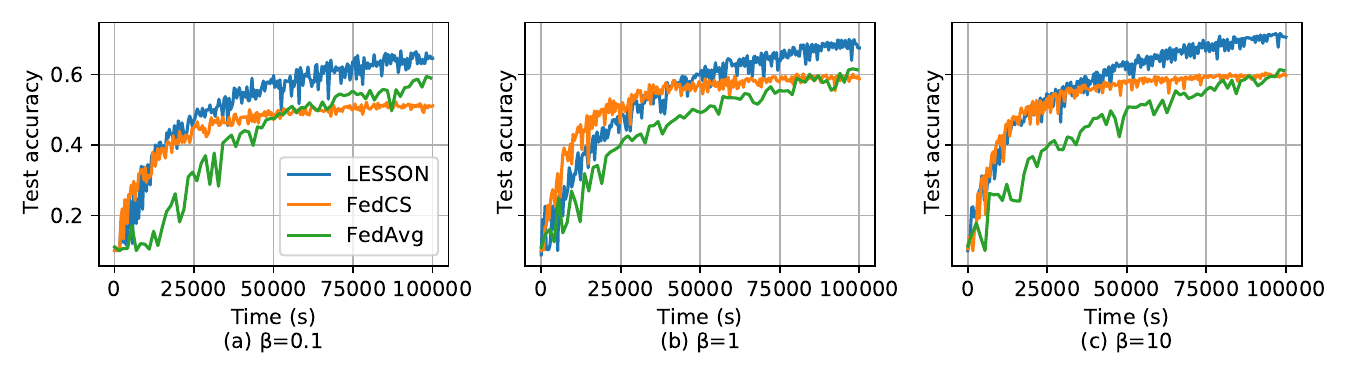} 
	\caption{Test accuracy over the time for CIFAR-10, where (a) $\beta=0.1$, (b) $\beta=1.0$, and (c) $\beta=10.0$.}
	\label{fig:var_beta_time} 
\end{figure*}

We further investigate how the data sample distribution affects the performance of the algorithms based on CIFAR-10. As mentioned before, $\beta$ is used to change the data sample distribution, i.e., a larger $\beta$ implies a more balanced data partition among the labels in a client or data sample distribution much closer to IID, and vice versa. Assume $\tau=20$ seconds, and Fig. \ref{fig:var_beta_iteration} shows the test accuracy of different algorithms over the number of global iterations by selecting different values of $\beta$. From the figures, we can see that if the data sample distribution exhibits non-IID, i.e., $\beta=0.1$, FedAvg has higher test accuracy than LESSON and FedCS. As $\beta$ increases, i.e., the data sample distribution is getting closer to IID, the test accuracy gap between FedAvg and LESSON is getting smaller, while the test accuracy of FedCS remains unchanged. On the other hand, Fig. \ref{fig:var_beta_time} shows the test accuracy of different algorithms over the time by selecting different values of $\beta$. From the figure, we can see that LESSON achieves much faster convergence rate than FedAvg and FedCS under different values of $\beta$. Therefore, we conclude that LESSON achieves a faster convergence rate at the cost of reducing the model accuracy, especially when the data distribution exhibits non-IID.


\begin{figure*}[!htb]
	\centering	
	\includegraphics[width=\textwidth]{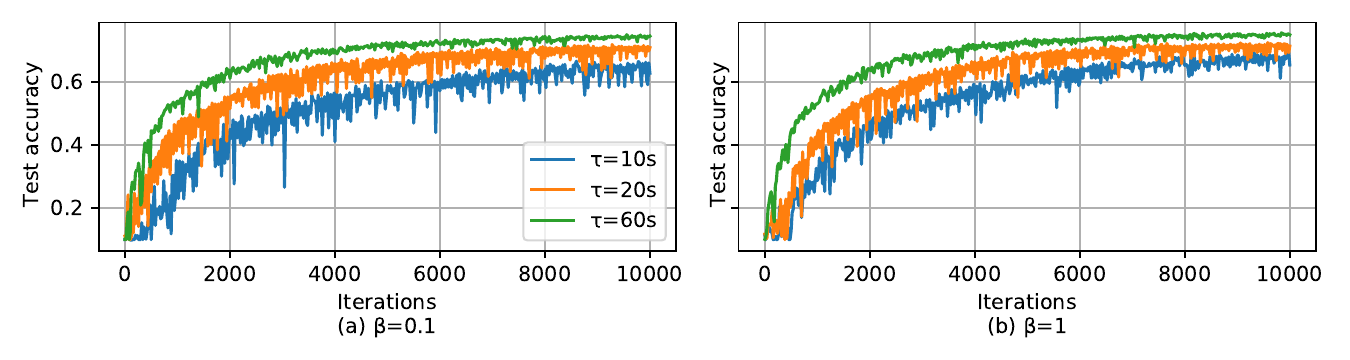} 
	\caption{Test accuracy over the number of global iterations for CIFAR-10, where (a) $\beta=0.1$ and (b) $\beta=1.0$.}
	\label{fig:var_deadline_iteration} 
\end{figure*}

\begin{figure*}[!htb]
	\centering	
	\includegraphics[width=\textwidth]{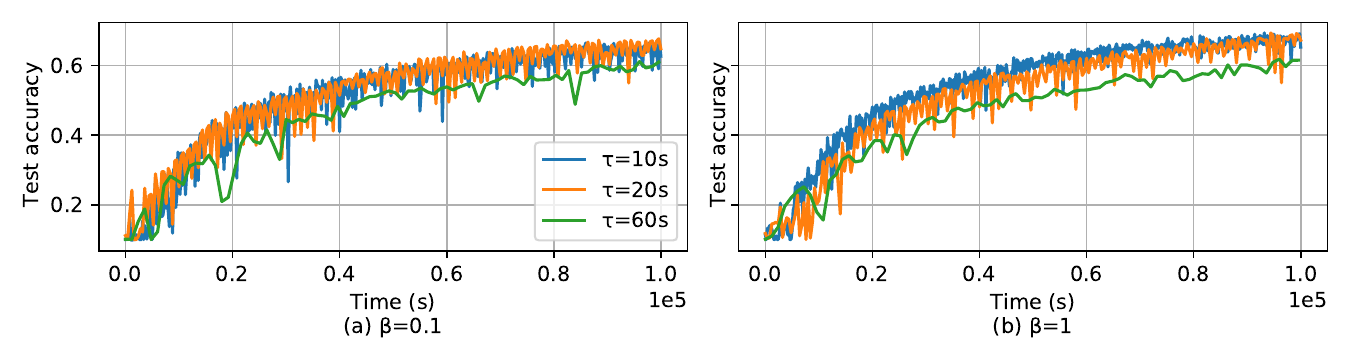} 
	\caption{Test accuracy over the time for CIFAR-10, where (a) $\beta=0.1$ and (b) $\beta=1.0$.}
	\label{fig:var_deadline_time} 
\end{figure*}
As mentioned in Section III.C, the deadline of a global iteration, i.e., $\tau$, is a very important parameter to adjust the performance of LESSON. Basically, if $\tau \rightarrow +\infty$, LESSON acts as FedAvg, and if $\tau \rightarrow 0$, LESSON acts as asynchronous FL. Fig. \ref{fig:var_deadline_iteration} shows the test accuracy of LESSON by having different values of $\tau$ and $\beta$ for CIFAR-10 over global iterations, where we can find that $\tau=60$ seconds incurs the highest test accuracy for both $\beta=0.1$ and $\beta=1$. This is because a larger $\tau$ 1) reduces the number of tiers in the system, thus alleviating the performance degradation caused by stale issue, and 2) increases the average number of clients in uploading their local models in a global iteration, which can mitigate the impact caused by non-IID. Also, we can see that the blue curve, i.e., $\tau=10$ seconds, is more sensitive to the change of $\beta$ than the other two curves. This is because as $\tau$ reduces, LESSON acts more like asynchronous FL, which has the convergence issue under non-IID. Figs. \ref{fig:var_deadline_time} show the test accuracy of LESSON by having different values of $\tau$ and $\beta$ for CIFAR-10 over the time. From Figs. \ref{fig:var_deadline_time}(a), we can find that $\tau=60$ exhibits the slowest convergence rate with respect to the time because a larger $\tau$ implies a longer latency of a global latency, i.e., $\tau=60$ runs the fewest global iterations than $\tau=10$ and $\tau=20$ within a time period. Therefore, we conclude that changing $\tau$ can adjust the tradeoff between the model accuracy and model convergence rate with respect to the time. A large $\tau$ can increase the model accuracy but reduce the model convergence rate, and vice versa. Meanwhile, as the data distribution is closer to IID (i.e., as $\beta$ increases), the difference of the convergence rate among the three algorithms increases as shown in Fig. \ref{fig:var_deadline_time}(b).    

\section{Conclusion}
In order to solve the data diversity and straggler issues in the synchronous FL while avoiding model divergence in the asynchronous FL, we propose a semi-synchronous client selection and model aggregation algorithm, i.e., LESSON, which allows all the clients to participate in the FL process but with different frequencies. Simulation results demonstrate that, in the IID setting, the model accuracy incurred by LESSON and FedAvg is similar but high than FedCS, and the convergence rate incurred by LESSON is the highest. In the non-IID setting, FedAvg achieves the highest model accuracy at the cost of lowering the convergence rate, and LESSON still incurs the highest convergence rate. Also, the other advantage of LESSON is that we can adjust the deadline $\tau$ in LESSON to change the tradeoff between the model accuracy and model convergence rate with respect to the time. In the future, we will investigate on how to dynamically adjust $\tau$ such that overall performance of LESSON can be maximized.   


\bibliographystyle{IEEEtran}
\bibliography{IEEEabrv,mybibfile}

\end{document}